\documentclass[preprint,3p,times]{elsarticle}
\usepackage[version=4]{mhchem}
\usepackage{siunitx}
\usepackage {subcaption}
\usepackage{placeins}
\usepackage{csquotes}

\usepackage{diagbox}
\usepackage[colorlinks,linkcolor=blue,anchorcolor=blue,citecolor=blue]{hyperref}
\biboptions{sort&compress}
\usepackage{threeparttable}
\usepackage{booktabs}

\DeclareSIUnit\bar{bar}
\DeclareSIUnit\elementarycharge{\text{\ensuremath{e}}}

\journal{NIM A}

\title{Measurement of the electron-hole pair creation energy in a 4H-SiC p-n diode}

\begin{document}
\begin{frontmatter}
    \author{Andreas Gsponer \corref{cor1}}
    \ead{andreas.gsponer@oeaw.ac.at}
    \author{Matthias Knopf}
    \author{Philipp Gaggl}
    \author{Jürgen Burin}
    \author{Simon Waid}
    \author{Thomas Bergauer}
    \cortext[cor1]{Corresponding author}

    \address{Institute of High Energy Physics, Austrian Academy of Sciences, Vienna, Austria}

    \begin{abstract}
        For 4H silicon carbide (4H-SiC), the values for the electron-hole pair creation energy $\epsilon_{\text{i}}$ published in the literature vary significantly.
        This work presents an experimental determination of $\epsilon_{\text{i}}$ using $50$ $\mu$m 4H-SiC p-n diodes designed for particle detection in high-energy physics.
        The detector response was measured for $\alpha$ particles between $4.2$ MeV and $5.6$ MeV for 4H-SiC and a silicon reference device.
        Different $\alpha$ energies were obtained by using multiple nuclides and varying the effective air gap between the $\alpha$ source and the detector.
        The energy deposited in the detectors was determined using a Monte Carlo simulation, taking into account the device cross-sections.
        A linear fit of the detector response to the deposited energy yields $\epsilon_{\text{i}} = (7.83 \pm 0.02)\;\text{eV}$, which agrees well with the most recent literature.
        For the 4H-SiC detectors, a linewidth of \SI{28}{\kilo\electronvolt} FWHM was achieved, corresponding to an energy resolution of \SI{0.5}{\percent}.
    \end{abstract}

\end{frontmatter}

\section{Introduction}
\label{sec:intro}
4H-Silicon carbide (4H-SiC) is gaining popularity as a material for  detector development. It offers multiple advantages over the ubiquitous silicon (Si) detectors currently in use.
The wide bandgap of 4H-SiC allows for high-temperature operation~\cite{nava_silicon_2008} and enables operation without cooling even after irradiation to high fluence (especially relevant for future high luminosity experiments)~\cite{de_napoli_sic_2022}.
The high breakdown voltage and carrier saturation velocity of 4H-SiC promises to enable enhanced timing performances compared to Si~\cite{Yang_2022, barletta_fast_2022}.

For the design of 4H-SiC detectors and their readout electronics, a proper understanding of 4H-SiC material properties is essential.
Among these are the ionization energy $\epsilon_{\text{i}}$ (energy needed to create an electron-hole pair) and the intrinsic energy resolution described by the Fano factor $\textit{F}$.
In the literature, there is still a considerable disagreement over their respective values, likely also due to the ongoing considerable improvements in material quality~\cite{de_napoli_sic_2022}.
For example, a review paper from 2023~\cite{sic_spectroscopy_72} uses $\epsilon_{\text{i}} = \SI{7.2}{\electronvolt}$, which disagrees with values in the range of \SIrange{7.6}{7.8}{\electronvolt} found in other publications.

In this paper, we thus provide an overview of 4H-SiC values for $\epsilon_{\text{i}}$ and $\textit{F}$ available in the literature. We further present measurements on SiC and Si-based particle detectors, both designed primarily for high-energy physics (HEP) applications, to determine $\epsilon_{\text{i}}$ of a SiC p-n diode.
Differences in the detector cross-section are accounted for using a precise GATE~\cite{GATE} simulation.
By controlling the air pressure during the measurement (from vacuum to ambient air pressure), a range of energies and distributions can be accessed using the same $\alpha$ source, presented in Section~\ref{sec:results}.
For extracting $\epsilon_{\text{i}}$, we compare the signals for SiC and Si and scale the ratio with the well-known ionization energy of the latter.
Finally, we assess and discuss the energy resolution of SiC detectors in Section~\ref{sec:spectrum}.

\subsection{Ionization Energy and Fano Factor}
The average energy a single primary particle has to expend to create one electron-hole pair on its passage through matter is referred to as the \emph{ionization energy} $\epsilon_{\text{i}}$.
It can be used to convert the incident energy of a primary particle to an amount of generated charge carrier pairs $n_\text{e-h}$, provided that the particle stops in the medium, thus transferring its total kinetic energy.
In an ideal detector, all events along a particle track are considered independent, and the total number of electron-hole pairs is equal to $n_\text{e-h} = E/\epsilon_{\text{i}}$, where $E$ is the total deposited energy.
The stopping of particles in matter is subject to energy straggling and exhibits statistical fluctuations in the number of individual processes and energy losses.
This process may be described by the Poisson statistic and leads to signal fluctuations on the order of $\sqrt{n_\text{e-h}}$.
However, due to the fact that the individual particle-particle interactions are not statistically independent~\cite{fano_original}, the energy resolution is improved by a factor $\sqrt{\textit{F}}$.
This \emph{Fano factor} $\textit{F}$ is a measure of the intrinsic energy resolution for a given material due to fluctuations in the primary signal formation process.
It is defined as the ratio of the observed variance $\sigma^2$ to the variance of the Poisson statistic $\sqrt{n_\text{e-h}}$
\begin{equation}
    \textit{F} = \frac{\text{observed variance}}{\text{Poisson variance}} = \frac{\sigma^2}{E/\epsilon_{\text{i}}}.
\end{equation}
With this, the standard deviation in terms of energy $\sigma_{\text{Fano}}=\sigma/\epsilon_{\text{i}}$ can be written in the usual form:
\begin{equation}
    \sigma_{\text{Fano}} = \sqrt{\epsilon_{\text{i}} \textit{F} E}
\end{equation}
In practice, however, there are also other effects contributing to the variation of the measured energy resolution $\sigma_{\text{meas.}}$, i.e.,
\begin{equation}
    \sigma_{\text{meas.}}^2 = \sigma_{\text{Fano}}^2 + \sigma_{\text{noise}}^2 + \sigma_{\text{absorbed}}^2
\end{equation}
The contribution $\sigma_{\text{noise}}$ corresponds to the noise of the readout electronics and $\sigma_{\text{absorbed}}$ to energy straggling caused by material that covers the sensitive volume of the detector~\cite{SiC_Fano}.
Losses due to the variation in charge collection can be minimized by using high-quality materials and adequate bias voltages.

The literature reports a wide range of ionization energies for 4H-SiC, with values ranging between \SI{5.05}{\electronvolt}~\cite{SiC_505_eV} and \SI{8.6}{\electronvolt} \cite{SiC_86_eV}.
A summary is shown in Table~\ref{tab:literature_values}.
\begin{table*}[htp]
    \centering
    \caption{Previously published values of $\epsilon_\text{i,SiC}$ for 4H-SiC. For measurements that compare to silicon, the value of $\epsilon_\text{i,Si}$ is indicated. The reference labeled \enquote{*} refers to the results obtained in this work.} \label{tab:literature_values}
    \smallskip
    \small
    \begin{tabular}{l|c|c|c|c|c|c}
        \hline
        Ref.               & Year & $\epsilon_{\text{i,SiC}}$ [\si{\electronvolt}] & Radiation            & Method      & Device Type & $\epsilon_{\text{i, Si}}$ [\si{\electronvolt}] \\
        \hline
        *                  & 2023 & $7.83 \pm 0.02$                                & $\alpha$             & Comp. to Si & Diode       & 3.62                                           \\
        \cite{Garcia_2013} & 2013 & $7.82 \pm 0.02$                                & $\alpha$             & Charge inj. & Schottky    & -
        \\
        \cite{SiC_Fano}    & 2013 & 7.28                                           & $\alpha$             & Charge inj. & Schottky    & -                                              \\
        \cite{SiC_505_eV}  & 2006 & 5.05                                           & $e^-$                & DC Current  & Diode       & -                                              \\
        \cite{SiC_778_eV}  & 2005 & $7.78 \pm 0.05$                                & $\alpha$             & Comp. to Si & Schottky    & 3.62                                           \\
        \cite{SiC_778_eV}  & 2005 & $7.79 \pm 0.09$                                & protons              & Comp. to Si & Schottky    & 3.64                                           \\
        \cite{SiC_76_eV}   & 2005 & 7.6                                            & X-Rays (\ce{^241Am}) & Comp. to Si & Diode       & 3.60                                           \\
        \cite{SiC_86_eV}   & 2004 & 8.6                                            & $\alpha$             & Comp. to Si & Diode       & 3.62                                           \\
        \cite{SiC_771_eV}  & 2004 & 7.71                                           & $\alpha$             & Comp. to Si & Schottky    & 3.62                                           \\
        \cite{SiC_78_eV}   & 2003 & 7.8                                            & X-rays (SEM)         & Comp. to Si & Diode       & 3.67                                           \\
        \hline
    \end{tabular}
\end{table*}
However, more recent measurements have converged in the range of \SIrange{7.6}{7.8}{\electronvolt}~\cite{SiC_Fano,SiC_771_eV,SiC_76_eV,SiC_78_eV}.
Different methods have been used, but most frequently, $\epsilon_{\text{i}}$ was determined via comparison to a silicon detector.
This method is described in more detail in the next section,~\ref{sec:intro:comparison}.
As the ionization energies obtained by this method depend directly on the $\epsilon_{\text{i}}$ of silicon, $\epsilon_{\text{i,Si}}$, the values used in each publication are listed in the table as well.
\begin{table*}[b]
    \centering
    \caption{Values of $F_\text{SiC}$ for 4H-SiC used in published literature.}\label{tab:literature_values:fano}
    \smallskip
    \small
    \begin{tabular}{l|c|c|c|c|c}
        \hline
        Ref.               & Year & $\textit{F}$ & Radiation & Method                        & Device Type \\
        \hline
        \cite{SiC_Fano}    & 2013 & 0.128        & $\alpha$  & Klein model~\cite{Klein_1968} & Schottky    \\
        \cite{SiC_Fano_01} & 2011 & 0.1          & X-rays    & Calculated from linewidth     & Schottky    \\
        \cite{SiC_76_eV}   & 2005 & $\leq 0.04$  & X-rays    & Upper limit from linewidth    & Diode       \\
        \cite{SiC_78_eV}   & 2003 & 0.12         & -         & Hypothesized                  & -           \\
        \hline
    \end{tabular}
\end{table*}
Due to the difficulty of measuring the Fano factor, only a few published values are available for 4H-SiC. An overview is given in Table~\ref{tab:literature_values:fano}.
Estimations have been made from the Fano factor of silicon~\cite{SiC_78_eV}, and using Klein's model~\cite{Klein_1968, SiC_Fano}.
Experimental data is available from X-ray measurements, with one publication observing no significant statistical broadening of the energy resolution (and estimating $F_{\text{SiC}}<\num{0.04}$~\cite{SiC_76_eV}), and another finding $F_{\text{SiC}} = \num{0.100}$~\cite{SiC_Fano_01}.

\subsection{Comparison Method with Silicon}
\label{sec:intro:comparison}
To determine $\epsilon_{\text{i}}$ based on a comparison to Si, a spectroscopic measurement is performed in the same experimental setup for a Si and a 4H-SiC detector.
In most cases, $\alpha$ sources are used for this purpose, as $\alpha$ particles are quickly absorbed even in thin detectors (with a range of about \SI{26}{\micro\meter} for \ce{^241Am} in Si~\cite{NIST_Astar}).
In an ideal experimental setup, the ionization energy can then be calculated from the ratio of the obtained signals ($\mu_\text{Si}$ and $\mu_\text{SiC}$), which is directly proportional to the number of generated charge carriers $n_\text{e-h}$:
\begin{equation}
    \label{eq:ei}
    \frac{\mu_\text{Si}}{\mu_\text{SiC}} = \frac{E_\alpha \epsilon_\text{i,SiC}}{E_\alpha \epsilon_\text{i,Si}} \Rightarrow
    \epsilon_\text{i,SiC}=\frac{\mu_\text{Si}}{\mu_\text{SiC}}\cdot \epsilon_\text{i,Si}.
\end{equation}
Here, $\epsilon_\text{i,Si}$ and $\epsilon_\text{i,SiC}$ denote the ionization energies of Si and SiC, respectively, and $E_\alpha$ is the energy deposited by the $\alpha$ particles.
The main advantage of this method is that no absolute charge calibration of the electronics is necessary~\cite{SiC_Fano}.

However, there are a couple of caveats:
This method assumes a charge collection efficiency (CCE) of \SI{100}{\percent} and an identical amplifier response for both detectors.
The former can be achieved by operating the detector with voltages above the full depletion voltage, while the latter can be minimized by using devices with similar capacitance or introducing a correction factor~\cite{SiC_778_eV}.
A major source of uncertainties is the energy loss in the detector layers above the sensitive volume (passivation and metal layers).
If there is a difference in these layers among devices, this needs to be considered as well, using stopping power calculations~\cite{SiC_778_eV} or Monte Carlo simulations~\cite{SiC_Fano}.
\section{Materials and Methods}
\label{sec:materials}
\begin{figure*}[b]
    \begin{subfigure}[b]{0.28\textwidth}
        \centering
        \includegraphics[width=.99\textwidth]{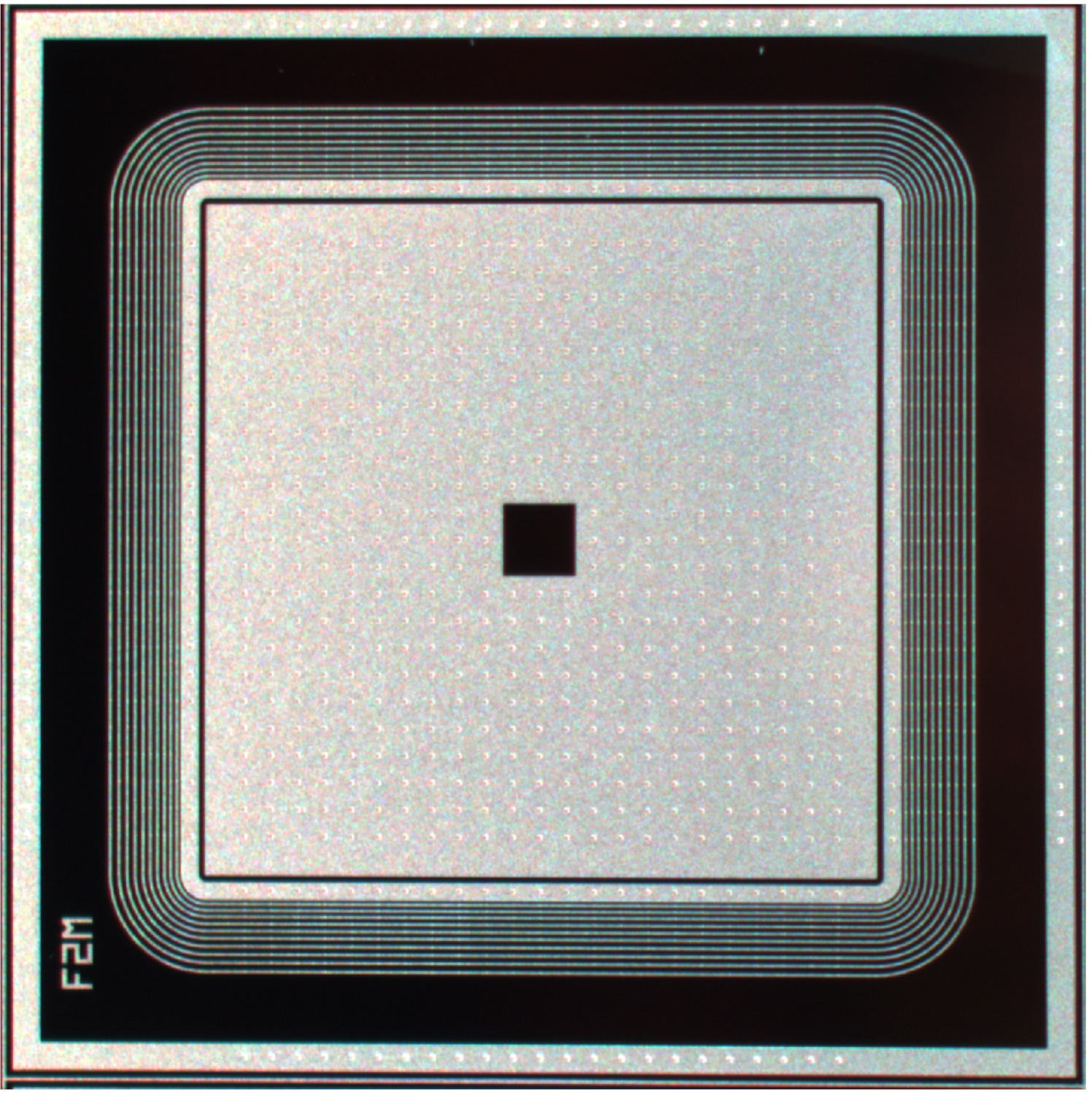}
        \caption{Silicon detector}
        \label{fig:materials:detectors:Si_W4}
    \end{subfigure}
    \hfill
    \begin{subfigure}[b]{0.28\textwidth}
        \centering
        \includegraphics[width=.99\textwidth]{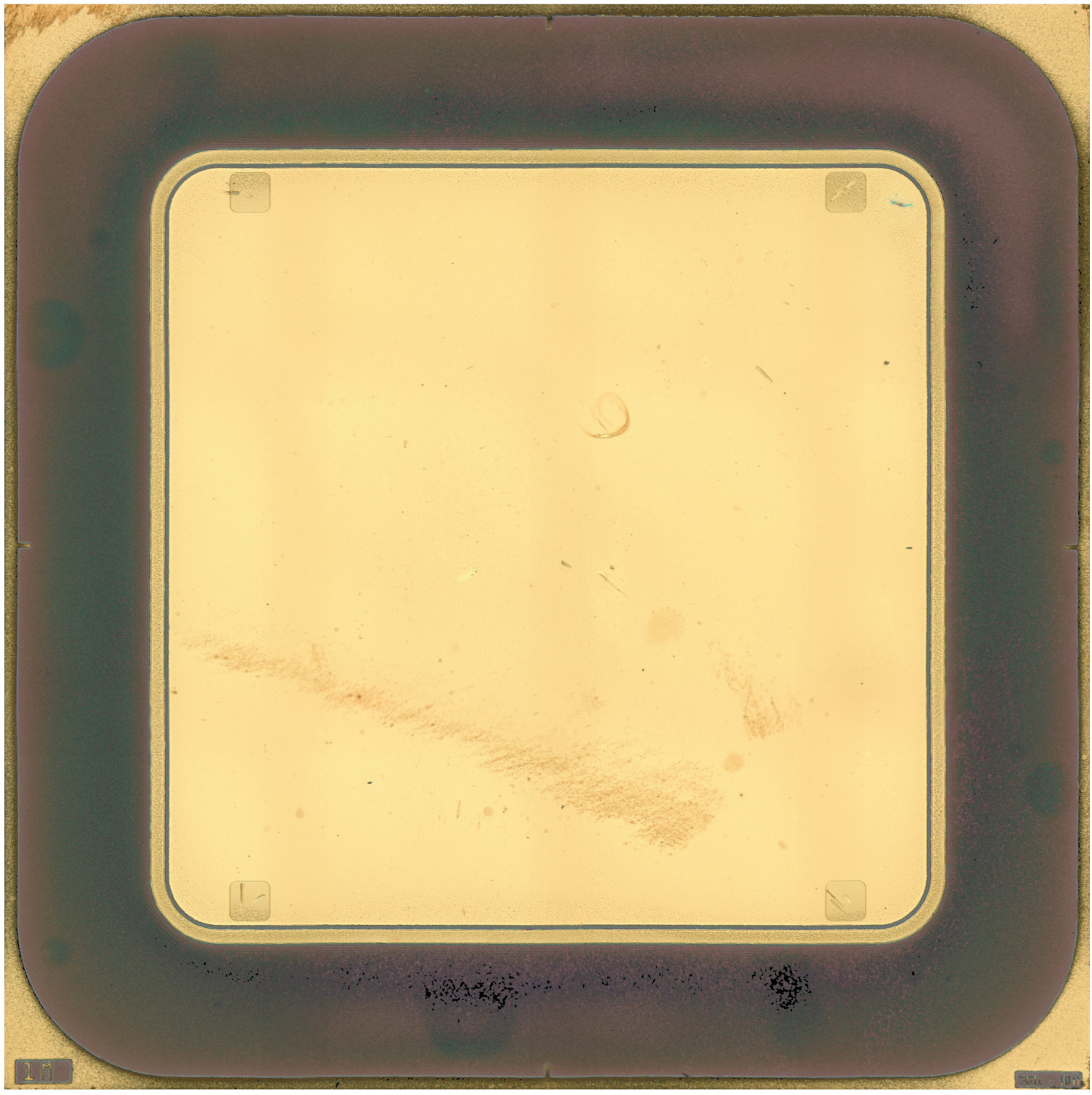}
        \caption{SiC detector with metallization}
        \label{fig:materials:detectors:SiC_1M}
    \end{subfigure}
    \hfill
    \begin{subfigure}[b]{0.28\textwidth}
        \centering
        \includegraphics[width=.99\textwidth]{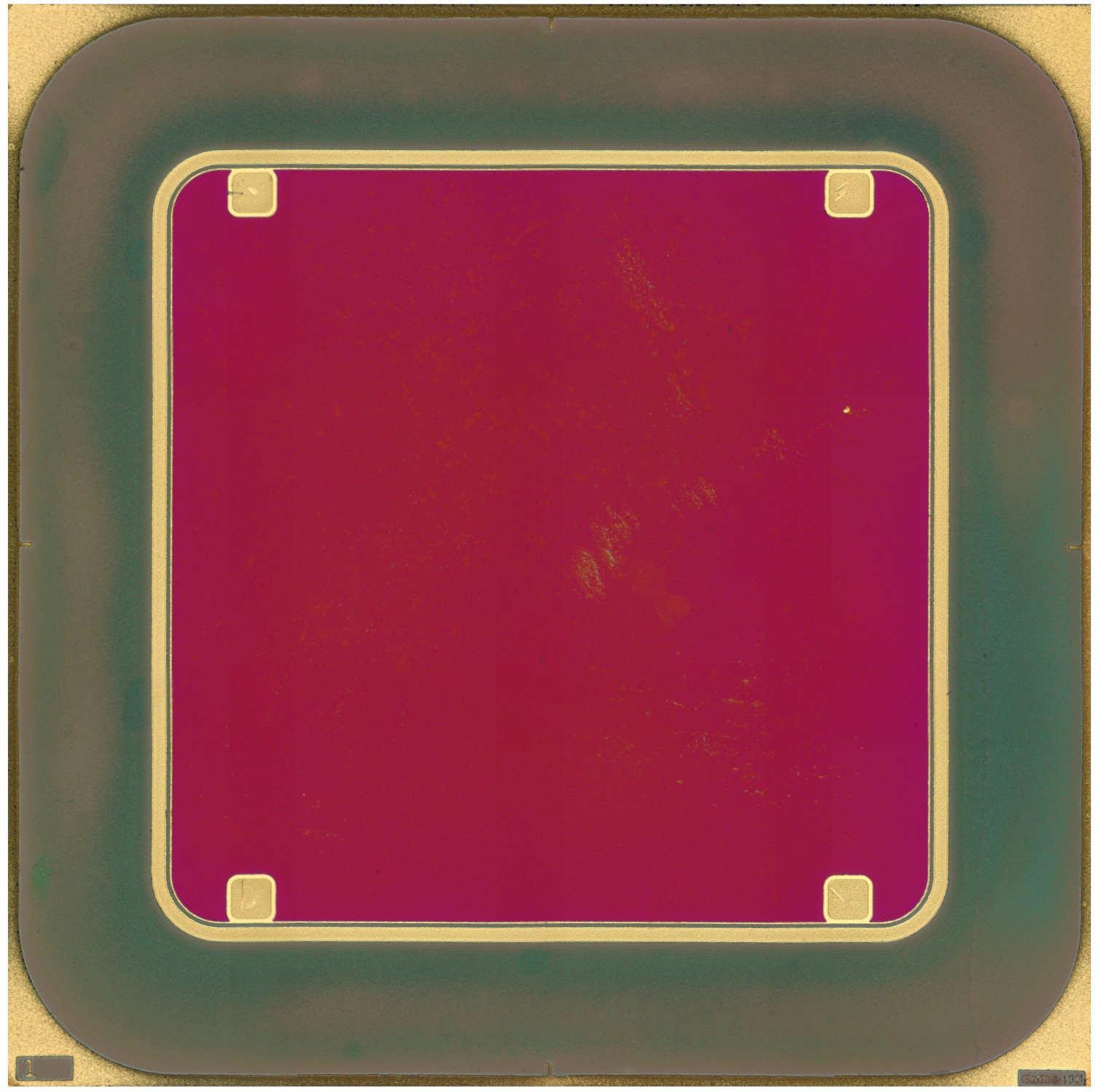}
        \caption{SiC detector without metallization}
        \label{fig:materials:detectors:SiC_1}
    \end{subfigure}
    \centering
    \caption{Microscopy images of Si and SiC detectors used in this work.}
    \label{fig:materials:detectors}
\end{figure*}

Two 4H-SiC detectors and a reference Si diode were used for the measurements, depicted in Figure~\ref{fig:materials:detectors}.
The employed SiC detectors were planar p-n diodes developed and manufactured by IMB-CNM-CSIC~\cite{CNM_SiC_Diode}, with two different types: One with metal covering the entire active area and another where the active area is only covered by the passivation layer.
Detectors from this production have previously been investigated using UV-TCT~\cite{gaggl_charge_2022}, $\alpha$ particles~\cite{gaggl_performance_2023}, and proton beams~\cite{christanell_4h-silicon_2022, gsponer_iworid_2023} before and after irradiation with neutrons.
These detectors featured a $3 \times 3$~\si{\milli\meter^2} large and  \SI{50}{\micro\meter} thick epitaxial layer with a resistivity of \SI{20}{\ohm\centi\meter} on a 4H-SiC substrate.
The metal layer of the detectors used titanium, aluminum, and nickel with a total thickness of \SI{1020}{\nano\meter}, while the passivation layer was composed of \ce{SiO_2} and \ce{Si_3N_4}~\cite{CNM_SiC_Diode, CNM_SiC_Diode2}.
For the sensor without metallization, metal is only present at the bond pads.
The thickness of the layers has been measured using transmission electron microscopy (TEM).
It corresponds to an equivalent aluminum thickness of \SI{1.72}{\micro\meter} for the detector with metal and \SI{0.60}{\micro\meter} for the detector without.
Full depletion was attained at \SI{300}{\volt}.
However, for the $\alpha$ source, a charge collection efficiency of \SI{100}{\percent} was already reached at \SI{100}{\volt}, due to the low penetration depth.\\
The silicon device was a planar n-type Si diode from the CMS production~\cite{cms_si_sensors}, shown in Figure~\ref{fig:materials:detectors:Si_W4}.
This diode had an active thickness of \SI{300}{\micro\meter} and a measured full depletion voltage of \SI{180}{\volt}.
The metal layer consisted of only aluminum (albeit thicker than for the SiC samples), with a passivation layer very similar to the 4H-SiC devices, for an equivalent aluminum thickness of \SI{2.23}{\micro\meter}.

When $\alpha$-particles travel through the air, they scatter and quickly lose energy (with an average range in the order of \si{\centi\meter}).
Small changes in the source-detector distance can lead to significant uncertainties if the experimental geometry is not reproducible.
In order to eliminate this effect, measurements can be performed in a vacuum.
In this work, however, we performed measurements in a vacuum and at different air pressures to obtain results for multiple $\alpha$-particle energies.
This additionally allows us to verify the simulation results of the energy lost in the passivation and metal layers.
In order to perform measurements at different air pressures, a Pfeiffer EVR116 valve and a Pfeiffer RPT200 pressure gauge were used to actively control the pressure inside the setup using a PID loop.
For the measurements assessing the energy resolution of the detectors, the maximum achievable vacuum of \SI{0.3}{\milli\bar} was used.

The Si and SiC detectors were operated using a reverse bias of \SI{400}{\volt} to ensure full depletion and, thus, a \SI{100}{\percent} charge collection efficiency.
All measurements were performed at room temperature (\SI{25}{\degree}).
As an $\alpha$-source, an Eckert \& Ziegler QCRB2500 spectrometric mixed nuclide source was used.
The radionuclides present in the source were \ce{^239Pu}, \ce{^241Am}, and \ce{^244Cm}, with an activity of \SI{1}{kBq} per isotope.
The main decay energies of the employed sources are almost equidistant, at \SI{5.157}{\mega\electronvolt}, \SI{5.483}{\mega\electronvolt}, and \SI{5.805}{\mega\electronvolt} ~\cite{NuDat3}.
The radionuclides were concentrated on a \SI{7}{mm} disk with a thickness of about \SI{1}{\micro\meter}, located behind a steel collimator.
The detectors were mounted and wire-bonded on passive ceramic PCBs.
A custom-made holder was designed, which allows for a constant source-detector distance of \SI{6}{mm}.
The entire setup was wrapped with aluminum foil and situated inside a stainless steel vacuum chamber to ensure RF shielding.

The spectroscopic measurements were performed with two readout chains.
The first readout chain, used for the pressure scan and determination of the SiC ionization energy (Section~\ref{sec:results}) was a Cividec Cx-L shaping charge-sensitive amplifier (CSA).
The histogramming of the CSA pulse heights was performed by a Rohde~\&~Schwarz RTO6 oscilloscope in HD mode (\SI{16}{\bit} resolution at an analog bandwidth of \SI{100}{\mega\hertz}).
The second readout chain, used to assess the energy resolution of the detectors (Section~\ref{sec:spectrum}), was composed of an Amptek CoolFET (A250CF based) CSA together with an Ortec 671 shaper and Ortec 928 multichannel analyzer.
A Keithley 2470 source meter was used to apply the reverse bias via a bias-T integrated into the respective CSAs.
\FloatBarrier
\section{Ionization Energy}
\label{sec:results}
\subsection{Simulation Results}
\label{sec:method:correction}
\begin{figure}[htp]
    \centering
    \includegraphics[width=.6\textwidth]{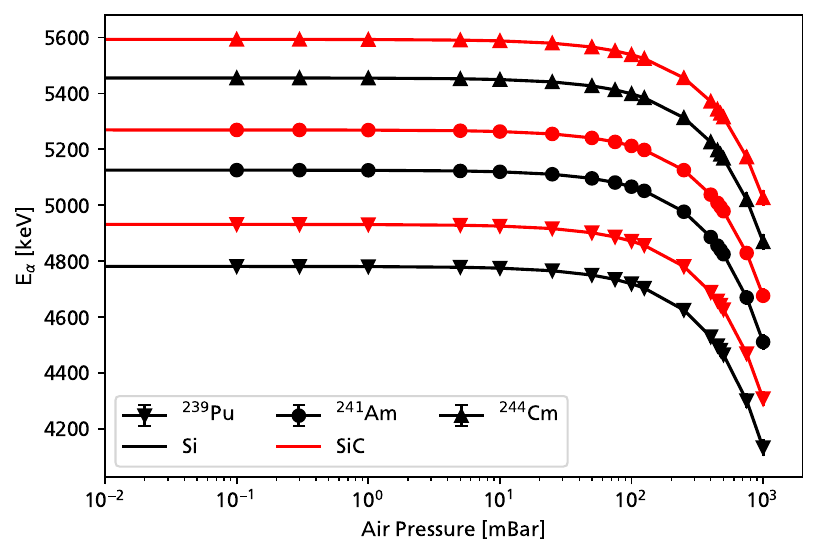}
    \caption{Energy deposited in detector $\text{E}_{\alpha}$ as a function of air pressure and nuclide for Si (black) and SiC with metallization (red).}
    \label{fig:gate}
\end{figure}

A GATE~\cite{GATE} simulation was used to determine the energy spectrum expected in the detectors (Si and SiC) as a function of the air pressure in the vacuum chamber.
For each nuclide and decay line, 5 million events were simulated, and the total energy deposited inside the detector was histogrammed.
Figure~\ref{fig:gate} shows the average energy per nuclide for both detector types.
As expected, the energy lost in the air gap increases with the air pressure.
An energy range of around \SI{500}{\kilo\electronvolt} can be accessed by tuning the pressure.
Due to the thicker metal layer of the Si device, $\alpha$-particles of about \SI{150}{\kilo\electronvolt} less energy reaches the sensitive volume than for the SiC device.

\FloatBarrier
\subsection{Measurement Results}
\begin{figure}[htp]
    \centering
    \includegraphics[width=.6\textwidth]{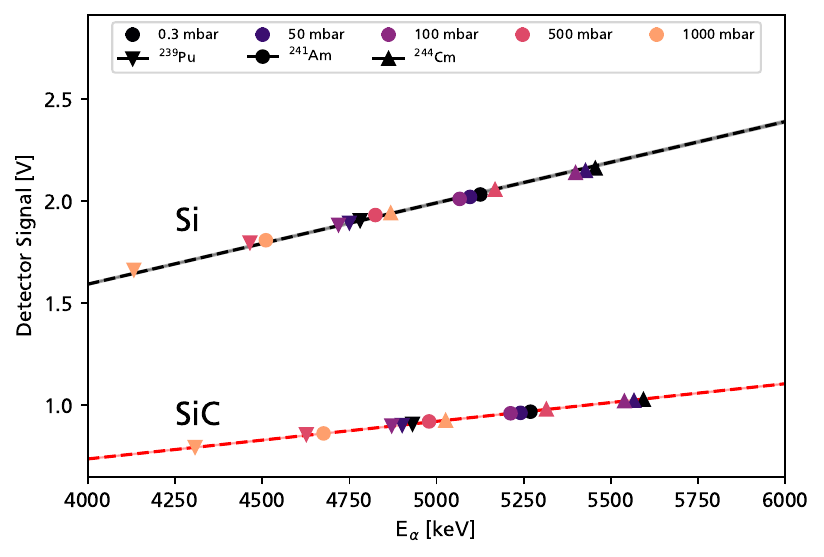}
    \caption{Detector signal versus energy in the detector. The color of the measured data points corresponds to the pressure inside of the vacuum chamber. For SiC, the detector with metallization (Figure~\ref{fig:materials:detectors:SiC_1M}) been used.}
    \label{fig:results}
\end{figure}

Measurements were performed in air at pressures between \SI{0.3}{\milli\bar} and \SI{1000}{\milli\bar}.
For each pressure setting, \num{25000} events were recorded for the Si and metallized SiC detectors.
Due to sample availability, no measurements with the SiC sample without metallization were performed.
As the individual decay energies of the nuclides could not be separated in the data (due to the straggling in air), one single fit was used for each nuclide.
The individual decay energies (and their probabilities) introduce an asymmetric energy distribution, so a skew-normal distribution~\cite{Azzalini_2013} was used for the fits.
The probability density function $\phi(x)$ of this distribution is given by
\begin{equation}
    \label{eq:skewnorm}
    \phi(x) = \frac{2}{\omega \sqrt{2\pi}} e^{-\frac{\left(x-\xi\right)^2}{2 \omega^2}} \int_{-\infty}^{\alpha\left(\frac{x-\xi}{\omega}\right)} \frac{1}{\sqrt{2\pi}} e^{-\frac{t^2}{2}} \mathrm{d}t,
\end{equation}
with $\xi$ being the location, $\omega$ the scale, and $\alpha$ the shape parameters respectively.
The mean $\mu$ and variance $\sigma^2$ of this distribution are given by
\begin{align}
    \mu = \xi + \omega \frac{\alpha}{\sqrt{1+\alpha^2}} \sqrt{\frac{2}{\pi}}, &  & \sigma^2 = \omega^2\left(1 - \frac{2\alpha^2}{\left(1+\alpha^2\right) \pi}\right).
\end{align}
Figure~\ref{fig:results} shows the fitted center positions plotted versus the energy deposited in the detector.
The simulated energy deposition corresponds closely to the measured detector signal.
Linear fits were performed for the Si and SiC data, resulting in a slope of $a_{\text{Si}} = \SI[separate-uncertainty=true]{0.398(1)}{\volt\per\mega\electronvolt}$ and $a_{\text{SiC}} = \SI[separate-uncertainty=true]{0.184(2)}{\volt\per\mega\electronvolt}$.
Taking the ratio of the slopes (compare with equation~\ref{eq:ei}) and using a silicon ionization energy $\epsilon_\text{i,Si}=$\SI{3.62}{\electronvolt}~\cite{Mazziotta_2008}, the ionization energy for 4H-SiC is obtained:
\begin{equation*}
    \epsilon_\text{i,SiC}=\SI[separate-uncertainty=true]{7.83(2)}{\electronvolt}
\end{equation*}
In this configuration, the RMS noise of the CSA was measured to be \SI{6.2}{\kilo\elementarycharge}, which equates to a FWHM of \SI{51}{\kilo\electronvolt} for the Si detector and \SI{117}{\kilo\electronvolt} for the SiC detector (a resolution of \SI{2}{\percent}).
The most likely origin of the CSA noise was due to insufficient filtering of RF noise injected by the SMU into the CSA bias-T.
The energy resolution of the SiC detectors is investigated more closely with lower noise electronics in the next section.
\FloatBarrier
\section{Energy Resolution}
\label{sec:spectrum}
\begin{figure*}[ht]
    \begin{subfigure}[b]{0.49\textwidth}
        \centering
        \includegraphics[width=.99\textwidth]{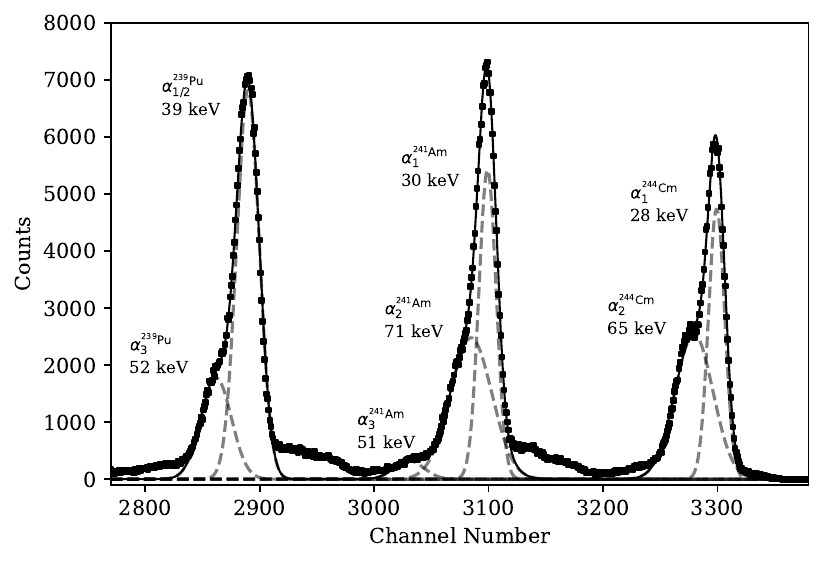}
        \caption{Measured Energy Spectrum for Silicon detector.}
        \label{fig:measurement:histo:Si_W4}
    \end{subfigure}
    \begin{subfigure}[b]{0.49\textwidth}
        \centering
        \includegraphics[width=.99\textwidth]{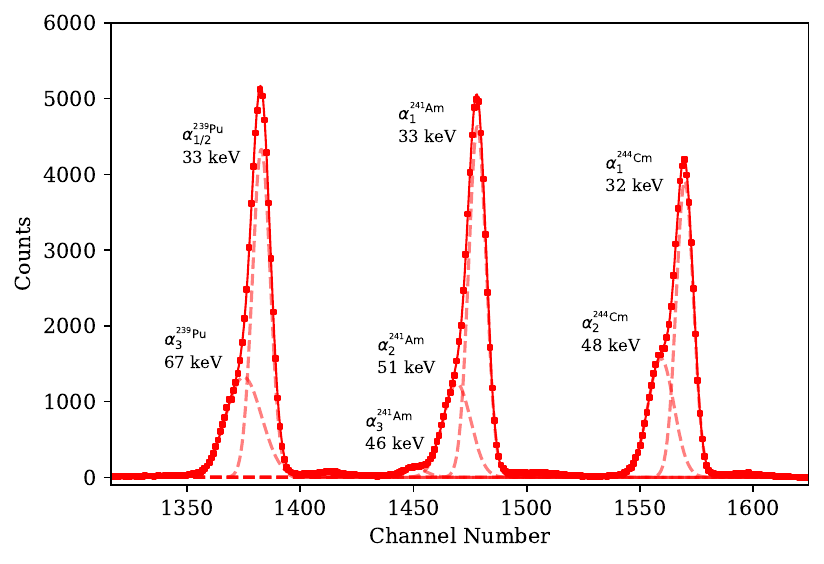}
        \caption{Measured Energy Spectrum for SiC detector with metallization.}
        \label{fig:mmeasurement:histo:SiC_1M}
    \end{subfigure}
    \hfill
    \vspace{1em}
    \newline
    \begin{subfigure}[b]{0.48\textwidth}
        \centering
        \includegraphics[width=.99\textwidth]{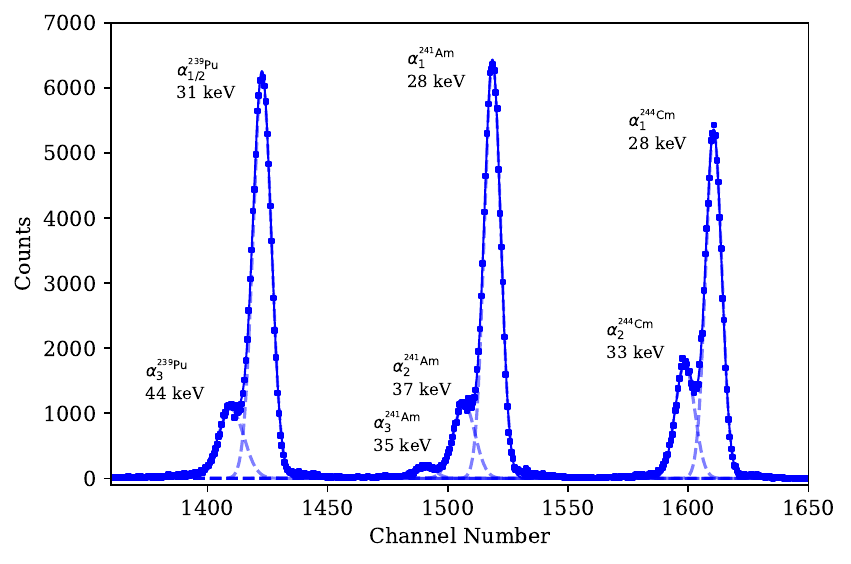}
        \caption{Measured Energy Spectrum for SiC detector without metallization.}
        \label{fig:measurement:histo:SiC_1}
    \end{subfigure}
    \centering
    \caption{Histograms of the measured $\alpha$ spectrum at a vacuum of \SI{0.3}{\milli\bar} for the Si (black) and SiC (red, blue) detectors. The Gaussian fits for each decay energy are indicated by dashed lines, with the sum shown using a solid line. For each energy, the fitted  FWHM resolution is annotated.}
    \label{fig:measurement:spectra}
\end{figure*}

In order to evaluate the spectroscopic performance of the detectors and to investigate the influence of the Fano factor, measurements using the CoolFET-based electronics chain were performed for the silicon and SiC detectors (with and without metallization).
Using this readout chain and an optimized Gaussian shaping of \SI{2}{\micro\second}, an RMS electron-noise-equivalent of \SI{560}{\elementarycharge} was obtained using a precision pulser while the Si and SiC detectors were connected and biased.
This corresponds to a FWHM resolution of \SI{5.2}{\kilo\electronvolt} for Si and \SI{10.3}{\kilo\electronvolt} for SiC.
For each detector, spectra were recorded for \num{8} hours at a vacuum of \SI{0.3}{\milli\bar} and are shown in Figure~\ref{fig:measurement:spectra}.
With the low noise CSA, the individual decay energies of the nuclides (\ce{^239Pu}, \ce{^241Am}, and \ce{^244Cm}) could be distinguished and fitted using a mixture of Gaussians.
For \ce{^239Pu}, the two most prominent decay energies are only \SI{12.7}{\kilo\electronvolt} apart, and the fit was performed using a single Gaussian for both energies.

Figure~\ref{fig:measurement:histo:Si_W4} shows the obtained spectrum for the Si detector with a FWHM resolution of around \SI{30}{\kilo\electronvolt} for the most probable decay energies.
For the decay energies with lower counts, the energy resolution is degraded and a large background signal is observed.
This background is thought to be caused by non-uniformities in the metal covering the Si detector (see Figure~\ref{fig:materials:detectors:Si_W4}), which introduce differences in energy absorption.

For the silicon carbide detector with metallization (Figure~\ref{fig:mmeasurement:histo:SiC_1M}) a similar performance was obtained, with a \SI{33}{\kilo\electronvolt} FWHM resolution of the main decay energies.
Similar to the silicon detector, the energy resolution is decreased at lower counts.
A small asymmetric background is visible towards higher channel numbers.
As the channel number of this background is identical to the main peaks for the SiC detector without metal (see Figure~\ref{fig:measurement:histo:SiC_1}), this is thought to be related to particles impinging the area around the guard ring, where no metallization is present on the detector surface.

The SiC detector without metallization (Figure~\ref{fig:measurement:histo:SiC_1}) showed the best spectroscopic performance, achieving a FWHM energy resolution below \SI{30}{\kilo \electronvolt} (a resolution better than \SI{0.5}{\percent} for \ce{^244Cm}).
Additionally, the background counts are much lower than for the previous detectors, and a resolution of \SI{35}{\kilo\electronvolt} FWHM can still be obtained for the \SI{5388}{\kilo\electronvolt} decay of \ce{^{241}Am}, which has a decay ratio of just \SI{1.7}{\percent}.
To the author's knowledge, this is the best energy resolution for 4H-SiC p-n junction detectors reported in literature  to date.

Comparing the channel numbers between the Si and SiC detectors in this individual measurement, ionization energies of $\epsilon_\text{i,SiC, Metal}=\SI[separate-uncertainty=true]{7.88(9)}{\electronvolt}$ and $\epsilon_\text{i,SiC, No Metal}=\SI[separate-uncertainty=true]{7.79(7)}{\electronvolt}$ are extracted, consistent with the value obtained in the previous section.

For the detectors, the energy straggling contributions $\sigma_{\text{absorbed}}$ obtained from the GATE simulations are the following: \SI{15.3}{\kilo\electronvolt} for Si and \SI{12.0}{\kilo\electronvolt} / \SI{9.1}{\kilo\electronvolt} FWHM for the SiC detectors with / without metallization.
Together with the electronic noise contribution $\sigma_{\text{noise}}$ of \SI{10.3}{\kilo\electronvolt}, this is not sufficient to explain the observed energy resolution in the range of \SI{30}{\kilo\electronvolt} FWHM.
Two possible sources of the lower-than- expected energy resolution are hypothesized:\\
Inhomogeneities in the thickness of the passivation and metal layers can lead to differences in energy absorbed by the $\alpha$ particles and broaden the energy distribution.
For the metal layer, the inhomogeneities are related to areas around the guard ring, where metal is only partially present.
A collimator on top of the detector could be used to constrain the detector area to the uniform sensitive area in the center.
For the passivation layer, especially in the SiC detectors, the height of the passivation is not always perfectly uniform due to manufacturing variations (observed in TEM).
This could be further improved by either thinning the passivation or by polishing.\\
The second possible source of a decreased energy resolution is imperfect collimation of the $\alpha$-source.
If the incidence angle of the $\alpha$-particles on the sensor is not perfectly normal, this can lead to variations in the effective path length as well, again leading to stochastic fluctuations of the energy loss.
Investigations in optimized source collimator geometries are currently ongoing.

Assuming a Fano factor of $\textit{F} = 0.100$~\cite{SiC_Fano_01}, the expected Fano noise $\sigma_{\text{Fano}}$ is smaller than \SI{5}{\kilo\electronvolt} FWHM.
This implies that even with the best energy resolution currently obtained in SiC $\alpha$-spectroscopy (\SI{15.7}{\kilo\electronvolt} using a Schottky diode~\cite{Mandal_2020}), the Fano noise contributes only a negligible role.
As the Fano noise grows with the square root of the incident particle energy, it is much more prominent in the measurement of low energy x-rays and plays a less significant role in $\alpha$-spectroscopy.
Measurements using soft x-rays emitted by the nuclides present in the $\alpha$-source could be envisioned in the future to try and determine the Fano factor of 4H-SiC.
\section{Discussion}
\label{sec:discussion}
The ionization energy of 4H silicon carbide (4H-SiC), where significant discrepancies in the literature values still exist,  has been determined using a p-n diode designed for high-energy physics applications.
The measurements leveraged an $\alpha$ source and an air gap between the source and detector at different air pressures, resulting in a varying energy deposition in the detector.
By comparing the spectrum obtained for the 4H-SiC device with a silicon reference detector, an ionization energy of \(\epsilon_\text{i,SiC}=\SI[separate-uncertainty=true]{7.83(2)}{\electronvolt}\) was obtained.
This value compares well with the most recent results in the literature (see summary table~\ref{tab:literature_values}).
The spectroscopic performance of SiC detectors with and without a metallization layer has been assessed using a low noise charge-sensitive amplifier.
An energy resolution of up to \SI{28}{\kilo\electronvolt} FWHM (corresponding to \SI{0.5}{\percent}) has been obtained, allowing for a clear distinction of the individual $\alpha$-decay energies.
The limitations of the energy resolution are still not fully understood and could be improved in the future by a factor of \num{2}, as the contribution of electronics noise and scattering sum to only  \SI{13.7}{\kilo\electronvolt}.
Likely candidates are thought to be an imperfect collimation of the $\alpha$-source and nonuniformities in the thickness of the detector passivation layer.

\section*{Acknowledgments}
This project has received funding from the Austrian Research Promotion Agency FFG, grant number 883652. Production and development of the 4H-SiC samples was supported by the Spanish State Research Agency (AEI) and the European Regional Development Fund (ERDF), ref. RTC-2017-6369-3.

\bibliographystyle{elsarticle-num}
\bibliography{biblio.bib}
\end{document}